\documentclass[aps,floatfix]{revtex4}
\usepackage{graphics}
\usepackage{epsfig}
\usepackage{subfigure}
\usepackage{amsmath,amsfonts,amssymb,graphicx,times}

\begin{document}

\title{Scaling property and opinion model for interevent time of terrorism attack}

\author{Jun-Fang Zhu}\email{zjfbird@mail.ustc.edu.cn}
\author{Xiao-Pu Han}\email{hxp@mail.ustc.edu.cn}
\author{Bing-Hong Wang}\email{bhwang@ustc.edu.cn (corresponding author)}

\affiliation{Department of Modern Physics, University of the Science
and Technology of China, Hefei 230026, China}

\begin{abstract}
The interevent time of terrorism attack events is investigated by
empirical data and model analysis. Empirical evidence shows it
follows a scale-free property. In order to understand the dynamic
mechanism of such statistic feature, an opinion dynamic model with
memory effect is proposed on a two-dimension lattice network. The
model mainly highlights the role of individual social conformity and
self-affirmation psychology. An attack event occurs when the order
parameter of the system reaches a critical value. Ultimately, the
model reproduces the same statistical property as the empirical data
and gives a good understanding of terrorism attack.
\end{abstract}

\maketitle



\section{Introduction}
In recent years, terrorism attack events have occurred frequently.
It has attracted interests of many scientists such as historians,
politicians, physicists and so on. Many people might think terrorism
attack is random and unpredictable. But, indeed there are some
general, common statistic property, just like the result from other
fields where the non-poison distribution is captured for the
interevent time of the e-mail [1], surface mail[2], short message
lending [3], web browsing [4,5], and rating of movies [6] etc.

As early as 1948, Richardson found the number of casualty follows a
power-law distribution in interstate wars [7]. Recently the same
statistic property is revealed for the casualty numbers in the
global terrorism events [8]. Actually the terrorism events can be
thought as wars in war. However, so far, the fundamental reason of
terrorism events still remain unclear in view of its complexity and
diversity. At present, there are mainly three kinds of viewpoints to
explain the burst of terrorism events. The first kind is the
self-organized critical notion. At first, Cederman [9] gave a
possible interpretation for the finding of Richardson[7] by an
agent-based model. After 9/11 event, The analysis for Iraq,
Colombia, Afghanistan[10-12] displays the power law distribution
with a scaling exponent $\alpha=2.5$ which conforms with the non-G7
countries, the ones except the major industrialized nation: Canada,
France, Germany, Italy, Japan, the United Kingdom and the United
States, in old war [8]. Meanwhile the author proposed the
self-organized critical model of interaction among terrorist groups
who make decision of coalescence or fragmentation randomly or with
some one probability. This model produces the results coinciding
with the statistic data and give insight in term of the conception
of complex system. Further, it is generalized and perfected by
Clauset[13], and the solution of the steady-state behavior is
obtained analytically under the conditions of constant number of
terrorism-inclined individuals and the proportional relation between
the severity and the size of the attacking cell.
The second one proposed by Galam[14-17] is the terrorism model of
percolation theory based on individual passive supporters. In this
model one territory is under the terrorist threat if the density of
the passive supporters exceeds percolation threshold in this
territory. Further one clue is given to curb terrorism threat
without harming the passive supporters. The specific scheme is
increasing the value of terrorism percolation threshold by
decreasing the space dimension but not the number of nearest
neighbors. This interesting model describes the state of terrorism
and gives an orient to fight against terrorism. The third one is the
competition, selection viewpoint. Clauset et al. [18] found the
scale-free property of frequency-severity ditribution has evident
robustness on burst means and stability over time since 1968. they
have developed a toy model to explain this kind of behavior by the
mechanism of competition between states and the non-state actors
successfully. In another reference [19], from the strategic
selection, they illuminate the substitution and the competition in
the Israel-Palestine conflict is the reason whether an organization
resort to terrorism where the public standing is first thought a
origin of attack occurrence.

From the aspect of fighting against terrorism, besides the model of
percolation theory, the conditions of promoting a violence[20] are
also referetial. Lim et al.[20] think that a violence arises at
boundaries between groups with culture diffentiation when the group
size achieves a critical scale and point out the violence might be
prevented or minimized if appropriate boundaries is created for
current geocultural regions. However that is not indisputable[27].
Recently, the variation of the interevent time over time have been
concerned, and the research of Clauset et al.[21] shows the
organizational growth leads to the decrease of the interevent time.

In this paper we focus on the interevent time of terrorism events in
Iraq and Afghanistan from 2003 to 2007. It is found to obey Zipf's
law corresponding with the power-law distribution with scaling
exponents $\alpha=2.43$ and $\alpha=2.35$ for Iraq and Afghanistan
respectively. Considering the importance of memory[22-24], we
propose an opinion dynamic model with memory effect to understand
such a statistic property. Due to individual social conformity and
self-affirmation psychology, the public opinion is formed and varies
with time under the coaction of influence between individuals [25]
and individual history memory [26]. Ultimately, under some certain
social circumstance, an attack event occurs when the public
consensus reaches some degree. We have found that this model can
reproduce the same intervent time distribution of terrorism events.

\section{The empirical data}
Our data can be available from the database of MIPT
(http://www.terrorisminfo.mipt.org/incidentcalendar.asp), which
records the detailed information of terrorism attack events and
includes the domestic event after 1998. Since there are huge differences of terrorism events for different countries, our statistics are distinguished as different countries.

As well known, terrorism attack events frequently occur in Iraq and
Afghanistan due to the deep ethnic contradiction, intense religious
struggle and increasing anti-Western emotion. The patterns of
terrorism events occurrence in Iraq and Afghanistan are shown in
Fig. 1. Apparently the succession of events takes on a pattern with
long time inactivity. Then what statistic character does this kind
of activity pattern has? In order to make it clear, we study the
interevent time distribution of terrorism attack events in these two
countries from 2003 to 2007.

The interevent time obtained from everyday event number. It is
assumed that the events occur homogeneously and the interevent time
is the reciprocal of event number in this day if the event number is
more than one in a day. Otherwise the interevent time corresponds
the number of days between two consecutive events. Because the
diversity of interevent time is not enough, we draw the rank plots
in Fig. 2. Obviously there exist Zipf's laws. It suggests that the
time between two consecutive attack events is usually very short,
and the long interevent time is also can't be ignored. Indeed, a
Zipf's law can be consider as a cumulative distribution with a
power-law form. If the probability ditribution of interevent time
follows $p(\tau)=\tau^{-\alpha}$, the Zipf's exponent $\alpha'$
should satisfy the relation [28,29]
\begin{equation}
R^{-1/\alpha'}=\int_{R}^{\infty}p(\tau)d\tau  \propto
R^{-(\alpha-1)}.
\end{equation}
Here $R$ represents the rank. We find the Zipf's exponent
$\alpha'=0.70$ for Iraq and $\alpha'=0.73$ for Afghanistan. In term
of the above relation, we have $\alpha=1+1/\alpha'$. So the
interevent time distribution has power exponent $\alpha=2.43$ and
$\alpha=2.35$ for Iraq and Afghanistan respectively. For
Afghanistan, the smaller exponent means the interevent time is more
heterogenous than Iraq, namely, the occurrence of terrorism events
in Afghanistan shows more burstiness property.

\begin{figure}
\scalebox{0.9}[0.9]{\includegraphics{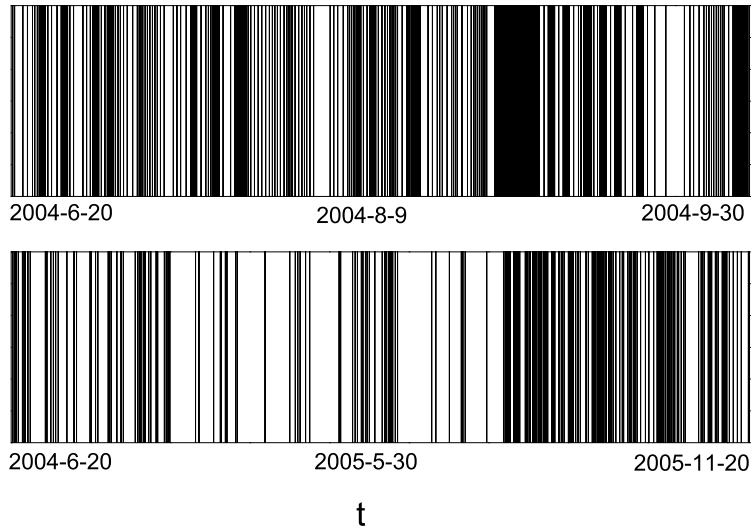}} \caption{The
succession of terrorism events in Iraq (upper panel) and Afghanistan
(lower panel). Each vertical line represents a single event.}
\end{figure}

\begin{figure}
\scalebox{1.2}[1.2]{\includegraphics{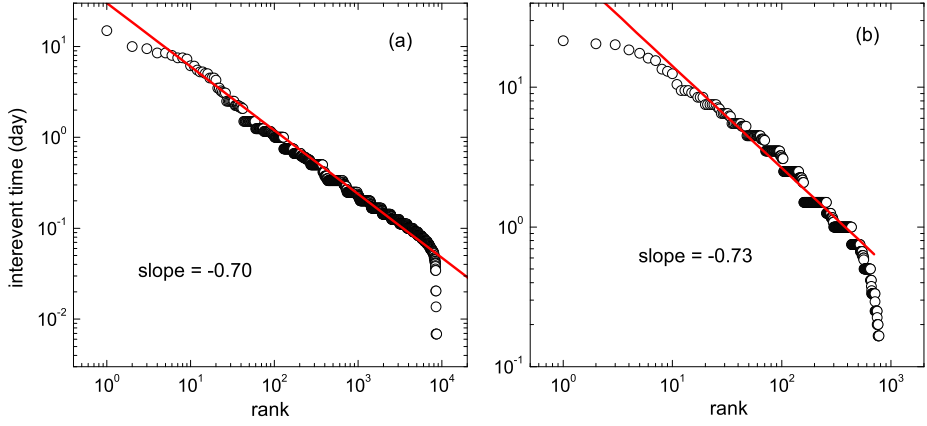}} \caption{The
interevent time distribution of terrorism attack events in Zipf's
plot for Iraq (a) and Afghanistan (b) from 2003 to 2007. The circles
represent the empirical data, and the solid line is the linear
fitting. }
\end{figure}

\section{the model}
To understand the underlying mechanism of the scaling property in
terrorism events, we introduce an opinion dynamic model with memory
effect to explain this statistic property. In the present agents
model, every node is connected with its four adjacent neighbors
inwardly and outwardly on a 2-dimension lattice network. Here nodes
and links represent the individuals in a terrorism social system and
the interaction between nodes respectively. Before an attack event
occurs every individual has own viewpoint, support or opposition
which are denoted by $\sigma_{i}=1$ and $\sigma_{i}=-1$. Individual
opinion is determined by two factors. One is the influence of its
adjacent neighbors, it is described by
\begin{equation}
W_1(\sigma_{i,t})=\sigma_{i,t-1}\sum\limits_{j=1}^{4}\sigma_{j,t-1}(j=1,2,3,4).
\end{equation}
the other is the individual history memory effect measured by
\begin{equation}
W_2(\sigma_{i,t})= \left\{
\begin{array}{cc}
1,
\sigma_{i,t-1}\sigma_{i,t-2}>0\\
0,\sigma_{i,t-1}\sigma_{i,t-2}<0.
\end{array}
\right.
\end{equation}
Individual opinion turnovers with time by the above factors in terms
of next rule: $W_1>0$ means individual has consistent opinion with
the majority of its adjacent neighbors and change his/her own
opinion with the probability $[exp(-aW_1)+exp(-bW_2)]/T$. $W_1<0$
corresponds with the opposite case. In this case history memory
effect dominates and individual changes his/her opinion with the
probability $exp[-bW_2]$. So the overturning probability is
\begin{equation}
P(\sigma_{i})=\left\{
\begin{array}{cc}
[exp(-aW_1)+exp(-bW_2)]/T,
W_1>0\\
exp(-bW_2), W_1< 0.
\end{array}
\right.
\end{equation}
Here the parameters $a$ and $b$ are the main factors indicating the
social conformity psychology and self-affirm psychology
respectively. $T \geq 2.0$ is an index to describe the social chaotic
degree. By this rule the system undergoes self-organization
evolution in the non-equilibrium state.

Now we measure the order degree of the public opinion by an order
parameter $m$ ($0\leq m \leq1$) on a lattice network with size
$L\times L$ and periodic boundary,
\begin{equation}
m=|\frac{1}{L^{2}}\sum\limits_{j=1}^{L^{2}}\sigma_{i}|,\sigma_{i}\in{(+1,-1)}
\end{equation}
Generally the total population of a terrorism social system is
approximately invariant. So let $L = 10$ and the initial states are
given randomly. No matter how complicated that the practical reasons
may be, we think an attack event is triggered when $m$ reaches a
critical value $m_{c}$. Next we investigate the influence of
different parameters on the interevent time statistic property by
simulation.

\section{Simulation and Analysis}

First, the social conformity will impact on the whole social order
degree and smaller $a$ indicates individual has larger willingness
to follow social public opinion. In Fig. 3 (a) the interevent time
distribution is plotted for different values of $a$ . It is shown
that the distribution transits to power-law style from
power-law-like style with a tail gradually when $a$ increases. For
power-law-like style at $a = 0.5$ a natural cutoff of tail is
executed, then the power exponent $\alpha$ is obtained by linear
fitting. Obviously, the power exponent becomes larger with the
increase of $a$, which means the distribution of terrorism events is
more inhomogeneous when individual inclines to follow public opinion
more easily. From the inset in Fig. 3 (a), one can see that $\alpha$
increases as $a$ increases and converges to a steady value when $a$
is large enough. This is because the effect of the public opinion is
so weak for large a that the dynamic evolution of terrorism attack
is hardly affected.

\begin{figure}
\scalebox{1.2}[1.2]{\includegraphics{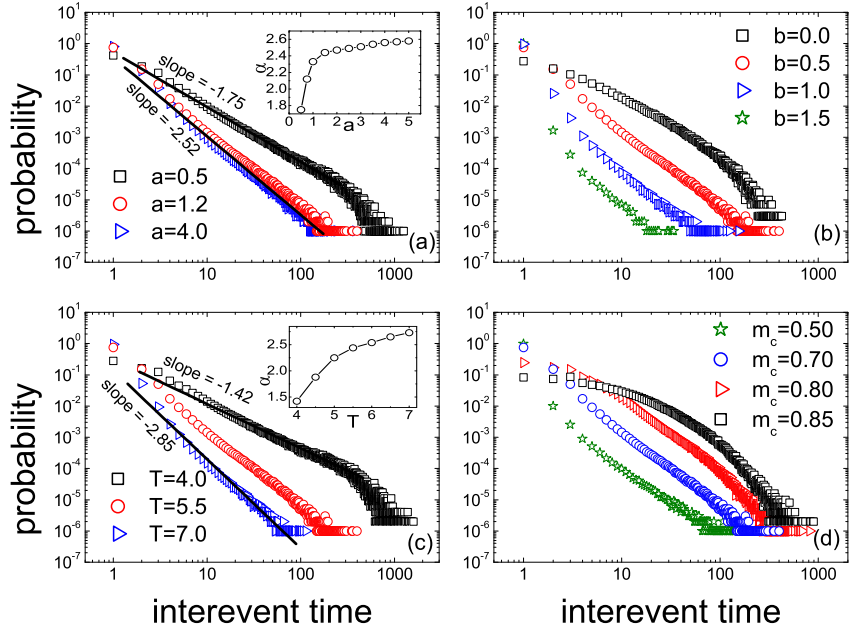}} \caption{(Color
online) The normalized distribution of interevent time by simulation
under the influence of one parameter among $a=1.2, b=0.5, T=5.5,
m_{c}=0.7$. (a) The distribution curves when the parameter $a$ is
changed. The two solid lines are the fitting of power-law with the
exponent $\alpha=1.75$ and $\alpha=2.52$ respectively. The top
inset: relationship between $a$ and $\alpha$. (b) The variation of
distribution curves when the paramter $b$ is changed. (c) the
distribution curves when the parameter $T$ is changed. Two curves
fit by two solid lines has scaling exponent $\alpha=1.42$ and
$\alpha=2.85$ respectively. The top inset: $\alpha$ as a function of
$T$. (d) The variation of distribution curves when the parameter
$m_{c}$ is changed. Each data is obtained by averaging over 100
independent runs.}
\end{figure}

In the evolution of terrorism events, self-affirmation psychology is
also very important for social order degree. Together with the
social conformity, they countermine wether an individual overturns
his/her opinion when individual opinion agrees with the major
opinion. However, in converse case, the self-affirmation play a
decisive role in individual opinion selection. At the moment,
individual will make a decision according to history selection in
memory. Now let us pay attention to the effect of self-affirmation
factor $b$ on time statistic property of terrorism event. As shown
in Fig. 3 (b), the distribution of interevent time exceeds a power
law when the self-affirmation is inadequate at large $b$ value. The
distribution curve becomes a power-law and then tends to a stretched
exponent style with the decrease of $b$ in that the strengthening of
self-affirmation reduces the time interval between two consecutive
terrorism events. Meanwhile it makes terrorism event with short
interevent time occur with smaller probability. Indeed social
conformity and self-affirmation are two factors who are
complementary to each other and mutual restraint to reach balance.

The third tunable parameter is $T$, which implies the chaotic degree
of a social circumstance. It will be effective as individual opinion
keeps consistent with the public opinion. Similar with Fig. 3 (a),
the increase of $T$ also leads to the increase of power exponent and
reduce of interevent time of terrorism events from Fig. 3 (c). But,
compared with the $a$, it is different that the large $T$ still has
obvious influence on power exponent from the inset.

The above three factors determine the order degree of public
opinion. We need to set a critical value of order parameter $m_{c}$
to judge whether a terrorism event burst. Fig. 3 (d) displays the
influence of different critical value $m_{c}$ on the curve style. It
is very easy to achieve consensus when $m_{c}$ is small. So the
terrorism event bursts frequently. Contrarily time interval of the
terrorism event becomes longer in a way. The curve style transits to
the stretched exponent because the proportion of the middling time
interval is prominent relatively.

From Fig. 3, one can find that the style of distribution curve is
decided by the self-affirmation psychology and the critical order
parameter. The social conformity and the chaotic degree of a social
circumstance decide the power exponent. So the specific power-law
style with some exponent will be got if only we choose appropriate
parameter values. According to the property above, we choose $a =
1.31, b = 0.50, T = 5.50, m_{c} = 0.70$ and $a = 1.20, b = 0.50, T =
5.50, m_{c} = 0.70$ to simulate the terrorism events in Iraq and
Afghanistan respectively. The simulation results are shown in Fig.
4. The present model generates the power-law distribution with
exponent $\alpha = 2.43$ and $\alpha = 2.35$ which accord with the
empirical data. It is noted that the strong self-affirmation and
weak social conformity are the significant character for Iraq.
Nevertheless these two factors are almost equivalent for
Afghanistan.

\begin{figure}
\scalebox{1.2}[1.2]{\includegraphics{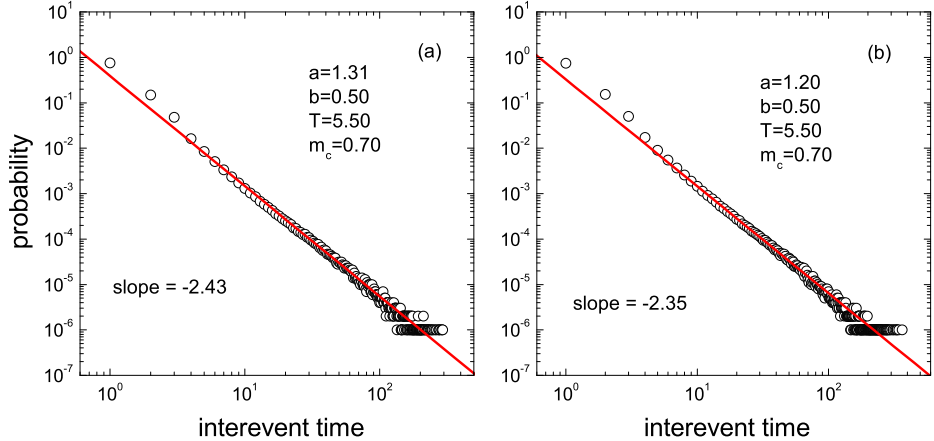}} \caption{The circles
represent simulation results for interevent time distribution of the
terrorism attack events in (a) Iraq and (b) Afghanistan and the
solid line is the linear fitting. Each data is obtained by averaging
over 100 independent runs.}
\end{figure}

\section{conclusion}

In conclusion, from real data we find the scale-free feature of
interevent time distribution for terrorism event in Iraq and
Afghanistan from 2003 to 2007. Here we consider the assumption that
the burst of a terrorism event is closely relative to the formation
of opinions. This formation process depends on not only the social
influence but also the individual memory. Previous terrorism models
have noted the former but ignored the later. So, to understand the
observed statistic property from empirical data, we proposed an
opinion dynamic model with memory effect in this paper. In the
model, the order degree of public opinion determines the burst of a
terrorism event. In certain social circumstance, the formation of
public opinion depends individual psychology character of social
conformity and self-affirmation. So individual psychology factor is
the crucial reason whether a terrorism burst in a given social
circumstance. This also alert us to poll is important in some wars.
Winning morale to strengthen the social conformity is a possible
means of reduce terrorism events. These results obtained by this
model are coincide with the reality intuitively and it can reproduce
the same power-law interevent time distribution of terrorism attack
as the empirical data in Iraq and Afghanistan. It confirms the
rationality of our assumption and provides a better understanding of
the terrorism attack. In addition, terrorism events can be treated
as a kind of collective behaviors of human. Our studies show that
the memory and social effect could be an origin of the power-law
properties in many collective behaviors of human.

The authors would like to thank Dr. Han-Xin Yang for helpful
conversation. This work is supported by the National Natural Science
Foundation of China (Grant No. 60744003, 10975126, and 70871082),
the Specialized Research Fund for the Doctoral Program of Higher
Education of China (Grant No. 20060358065),and the National Basic
Research Program of China (973 Program No.2006CB705500).

\end{document}